\documentclass[conference]{IEEEtran}
\IEEEoverridecommandlockouts
\usepackage{cite}
\usepackage{amsmath,amssymb,amsfonts}
\usepackage{algorithmic}
\usepackage{graphicx}
\usepackage{textcomp}
\usepackage{array}
\usepackage{xcolor}
\usepackage{multirow}
\usepackage{url}
\usepackage{atbegshi,picture}
\usepackage{lipsum}

\AtBeginShipout{\AtBeginShipoutUpperLeft{%
  \put( \paperwidth-3.66cm,-0.5cm){\makebox[0pt][r]{\framebox{Accepted for presentation in IEEE Biomedical Circuits and Systems Conference (BioCAS 2021).}}}%
}}

\def\BibTeX{{\rm B\kern-.05em{\sc i\kern-.025em b}\kern-.08em
    T\kern-.1667em\lower.7ex\hbox{E}\kern-.125emX}}
    
\makeatletter
    \newcommand{\linebreakand}{%
      \end{@IEEEauthorhalign}
      \hfill\mbox{}\par
      \mbox{}\hfill\begin{@IEEEauthorhalign}
    }
\makeatother
\begin{document}

\title{EMG Signal Classification Using Reflection Coefficients and Extreme Value Machine}

\author{
\IEEEauthorblockN{Reza Bagherian Azhiri }
\IEEEauthorblockA{\textit{Predictive Analytics and Technologies Lab, ME Dept.} \\
\textit{The University of Texas at Dallas}\\
Richardson, TX, USA \\
reza.azhiri@utdallas.edu}
\and
\IEEEauthorblockN{Mohammad Esmaeili}
\IEEEauthorblockA{\textit{Department of Electrical and Computer Engineering} \\
\textit{The University of Texas at Dallas}\\
Richardson, TX, USA \\
esmaeili@utdallas.edu}
\linebreakand
\IEEEauthorblockN{Mohsen Jafarzadeh}
\IEEEauthorblockA{\textit{El Pomar Institute for Innovation and Commercialization} \\
\textit{University of Colorado Colorado Springs}\\
Colorado Springs, CO, USA  \\
mjafarza@uccs.edu}
 \and
\IEEEauthorblockN{Mehrdad Nourani}
\IEEEauthorblockA{\textit{Predictive Analytics and Technologies Lab, ECE Dept.} \\
\textit{The University of Texas at Dallas}\\
Richardson, TX, USA \\
nourani@utdallas.edu}
}
\maketitle

\begin{abstract}
Electromyography is a promising approach
to the gesture recognition of humans if an efficient classifier with a high accuracy is available. In this paper, we propose to utilize Extreme Value Machine (EVM) as a  high performance algorithm for the classification of EMG signals. We employ reflection coefficients obtained from an Autoregressive (AR) model to train a set of classifiers. 
Our experimental results indicate that EVM  has better accuracy  in comparison to the conventional classifiers approved in the literature based on K-Nearest Neighbors (KNN) and Support Vector Machine (SVM).
\end{abstract}


\section{Introduction}

\begin{figure*}[t] 
\centering
\includegraphics[width=0.9\textwidth]{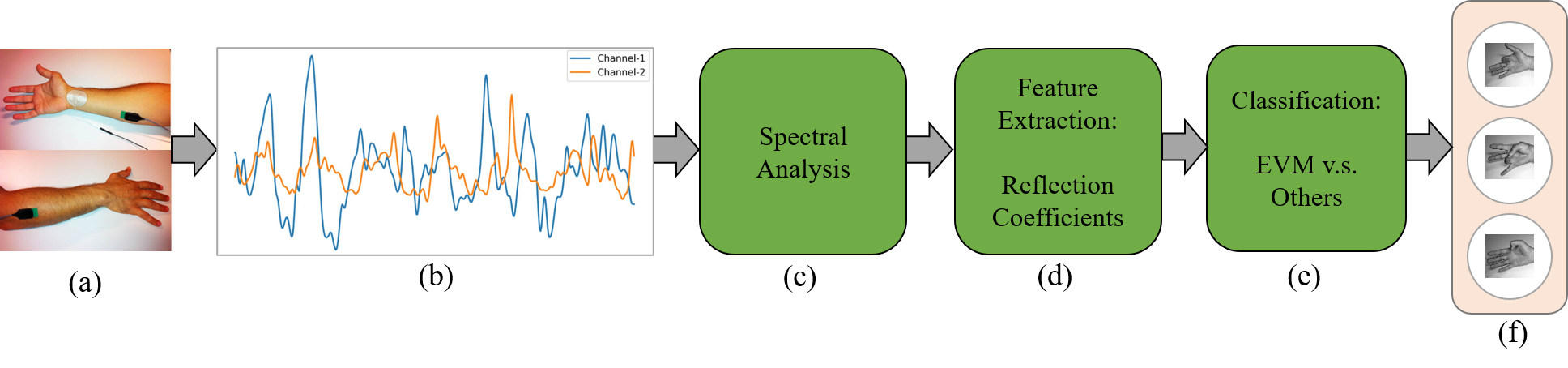}
\caption{Block diagram of our proposed model. (a) After muscle contraction, surface EMG electrodes read the (b) electric potential generated in muscle fibers. (c) Spectral analysis is implemented on obtained EMG signal. (d) Reflection coefficients is extracted as our features for classification. (d) For classification, the results of EVM are compared with other classifiers.}
\label{Teaser}
\end{figure*}

Limb amputation is a tragedy that could happen due to different reasons such as diabetes, vascular diseases, accidents or wars. Undoubtedly, mental and physical effects of missing a limb are severe and deprive individuals from normal daily activities. In the united states alone, it has been reported that the total number of upper limb amputees is about $41,000$ and is expected to grow by $131\%$ by $2050$~\cite{ziegler2008estimating}.
The monotonic  relation  between  finger  movement  and  that  of its  associated  electromyography (EMG) signal can be explored to control prosthetics. EMG is a technique to measure the electrical activity of nerve’s stimulation during the contraction. The data is collected by EMG electrodes placed on the surface of skin or intrusive into the skin in direct contact with the muscles~\cite{Riviere}. 

In fact, pattern recognition is a saddle point in dealing with EMG signals to classify these signals and relate the specified class to the corresponding command~\cite{Kamavuako, scheme2011electromyogram}. 
The feature extraction from EMG signals should be done such that the corresponding class with specific signal could be detected~\cite{Ning-Jiang}. Various feature extraction methods and classification algorithms have been employed by researchers in time and frequency domains~\cite{Knaflitz, Mohdkhan_review}. 
Then, various optimization methods such as Particle Swarm Optimization (PSO) and Genetic algorithm~\cite{2015novel, Lima2018genetic} can be applied to select a set of features with significant importance.

The number of extracted features depends on the number of EMG sensors and feature extraction strategy for each sensor. The dimensionality of features can thus be superabundant and may facilitate training of the system. 

In~\cite{Khushaba_towards}, two EMG electrodes are used for classification of ten classes of finger movements. The feature extraction includes Time Domain (TD) features, Autoregressive (AR) plus Hjorth features of EMG signals followed by Support Vector Machine (SVM) and k-nearest neighbor (KNN) classifiers for pattern recognition. In~\cite{Al-Timmemy}, twelve surface EMG electrodes are employed to extract features by combining time domain features and autoregressive. Finally, they employed SVM as a classifier of finger gesture. A notch filter and a Butterworth Band Pass Filter are utilized in~\cite{ref} to filter the signals. Then, the dataset and extracted features are normalized via Fractional Fourier Transform (FFT). Due to a large number of features, feature reduction is inevitable and is performed via t-test and windowing method. The final step in classification done by KNN~\cite{ref}. Our research group has previously approximated spectral analysis of EMG signals and extracted features using reflection coefficients~\cite{Heydarzade} and classified the results with SVM, without any feature reduction. 
It is shown in~\cite{Hachemi} that discrete wavelet transform followed by principal component analysis (PCA) and SVM could enhance the accuracy of classifications. Several other EMG classifiers have been applied using decision trees~\cite{Espinoza_DT_SVM}, random forest~\cite{Li_RF}, KNN~\cite{ref}, Naïve Bayes classifier~\cite{Praveen_naive}, and SVM~\cite{Heydarzade, gailey_svm}. Also, it seems using ensemble learning~\cite{forouzandeh2021presentation} and other neural network architectures such as a multilayer perceptron (MLP)~\cite{azhiri2021emg-based}, convolutional neural network (CNN)~\cite{jafarzadeh2019convolutional}, graph convolutional neural network (GCNN)~\cite{2020new}, and recurrent neural network (RNN)~\cite{Huang2019RNN}
can significantly improve the classification accuracy.

Block diagram of our proposed mode is demonstrated in Figure~\ref{Teaser}. To the best of our knowledge, this is the first time that the extreme value machine is proposed for classification of EMG signals. In this paper, the reflection coefficients of an autoregressive model are considered as the feature vector to serve as the input of the extreme value machine classifier.
We show the strength of this method by comparing the accuracy of extreme value machine against seven well accepted EMG classifiers proposed in the literature, i.e., K-Nearest Neighbor (KNN), Logistic Regression (LR), Support Vector Machine (SVM), Gaussian Naive Bayesian (GNB), Decision Tree (DT), Random Forest (RF), and XG Boost (XGB). 

This paper is organized as follows. The background of EVM is described in Section 2. Thereafter, in Section 3, the preprocessing technique is explained and discussed how feature extraction is done. In Section 4, the datasets characteristics are elaborated, and disparate classifiers are applied to the same dataset and results are compared. Finally, conclusion is drawn in the last section. 

\begin{figure*}[t]
\centering
\includegraphics[width=0.70\textwidth]{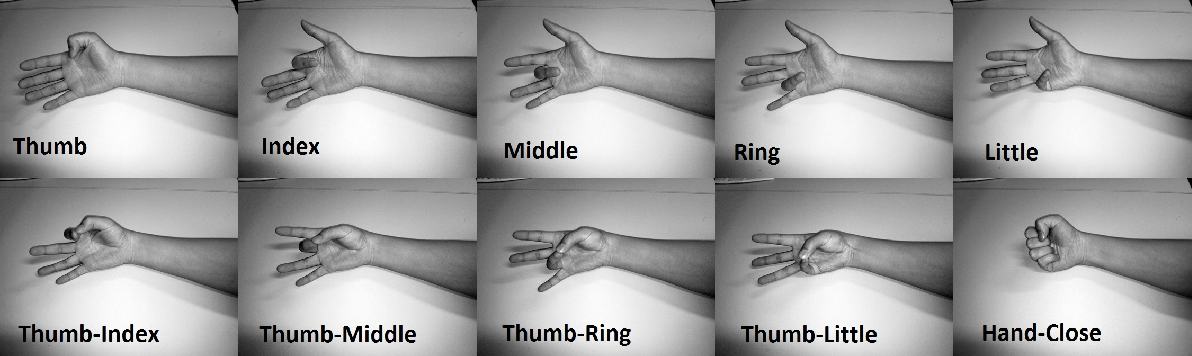}
\caption{Ten finger movements for individual and combined fingers\cite{Khushaba_towards}}
\label{gestures}
\end{figure*}

\begin{figure*}[t] 
\centering
\includegraphics[width=0.78\textwidth]{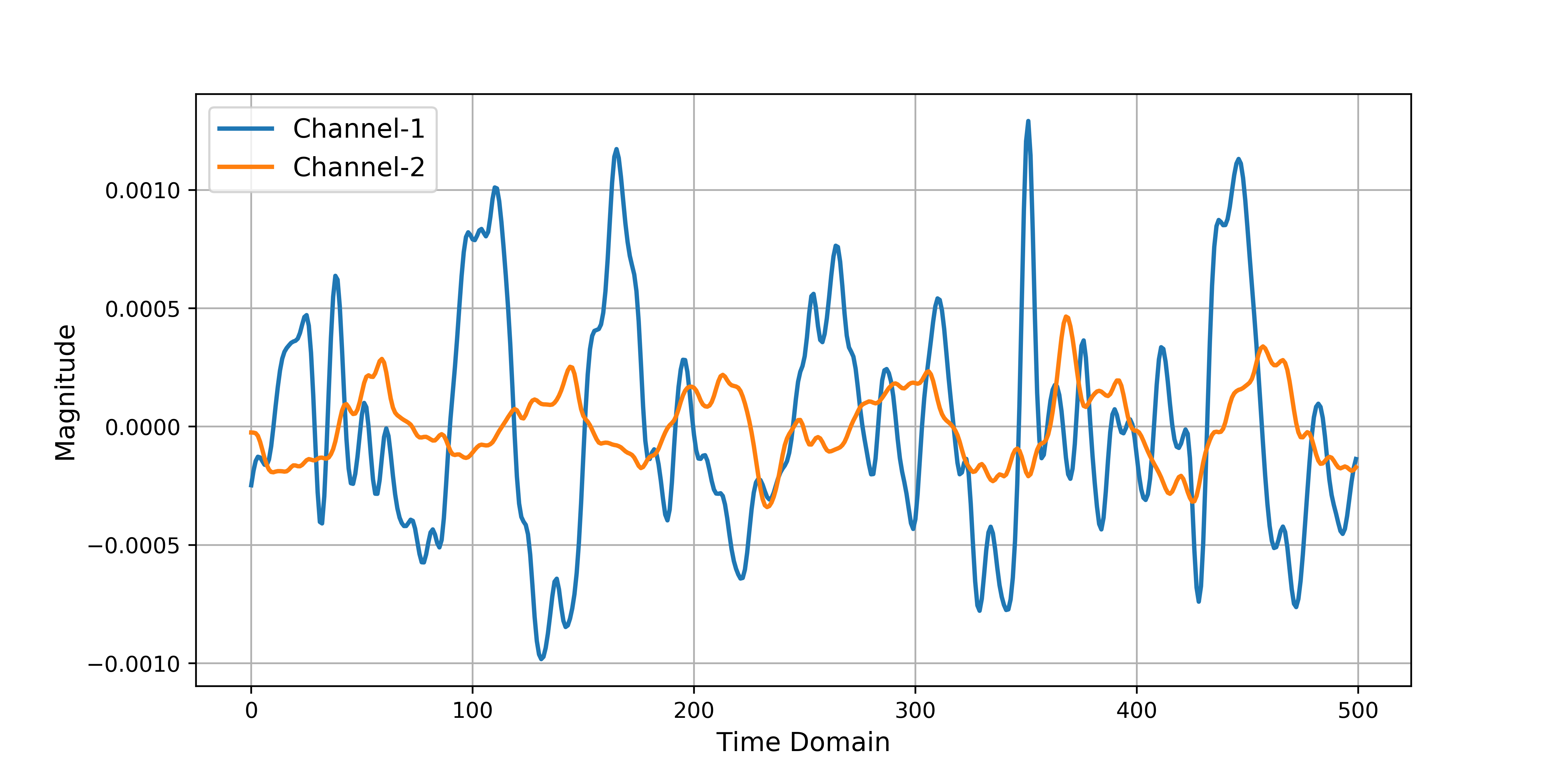}
\caption{EMG signal of combined finger movements (thumb-index)}
\label{EMG_signal}
\end{figure*}

\section{Extreme Value Machine}
\subsection{Background Theory}

Most of the classification algorithms do not consider the distribution of data in their learning process. Extreme Value Machine (EVM), derived from statistical Extreme Value Theory (EVT), is a distance-based kernel-free non-linear classifier which mainly focuses on the distribution of dataset especially at extreme values\cite{Rudd_EVM}. EVT is a reliable statistical approach to examine datasets that have too low or too high extreme values. In other words, EVT deals with stochastic behavior of rare observations that are unusually greater or smaller than other observations. This is important where the distribution of data at the tails of probability distribution is different from the ones at the center of  distribution. More specifically, the advantage of EVT is originated from the fact that because of using the asymptotic theory, the extremes (the tails of dataset distribution) are modeled effectively \cite{roberts2000extreme}.

Two different parametric methods have been proposed for finding the distribution of extreme values: peaks-over-threshold and block maxima~\cite{rouhani2018algorithms, beirlant2006statistics, castillo2012extreme}. In peaks-over-threshold approach, the concentration is only on the observations that exceed a pre-defined threshold. Threshold exceedance method converge to Generalized Pareto Distribution (GPD), a very popular extreme value model which could adequately extrapolate sufficiently high values located at the upper tail.  In block maxima approach, the observation period is divided into nonoverlap periods with the same size. Maximum observation at each period is considered for defining the distribution. The probability distribution of such observations is approximately generalized extreme value (GEV) distribution  \cite{jafarzadeh2021automatic}.

\subsection{Extreme Value Machine}

In EVM, the key factor is to define decision boundary based on probabilistic distribution of each class based on the extreme vector of corresponding class. The extreme vectors are commonly the points that determine the decision boundary of each class in a compact probabilistic form. The Weibull distribution is often used to compute the radial probability of inclusion of point considering the neighbor ones.
In fact, the aim of EVM is to select distributions of each class by extreme vectors and compact these probability distributions such that the decision boundary of classes could be determined. 
A Weibull distribution is fitted on margin distance of each extreme vector. Therefore, this distribution enables us to arrive to the decision boundary of each class could be defined with respect to its extreme vectors. Given a point to EVM, the probability of sample inclusion for the extreme vector of each class is calculated. If the inclusion probability of the sample point is more than a predefined threshold, the point is assigned to that class.  
For each point of the dataset like $x_i$, the prediction for each class is calculated by:

\begin{equation}
    \hat{P}(C_l|x) = \underset{k}{\textrm{max}} \textrm{W}_{l,k} (x;\mu_{l,k}, \sigma_{l,k}, \xi_{l,k})
\end{equation}
where $x$ is feature of input, $W_{l,k}(x)$ is probability density function of Weibull for  $k$th of  extreme vector of class $l$, $\xi$ is shape parameter of Weibull probability density functionm, $\mu$ is shift parameter of Weibull probability density functionm, and $\sigma$ is scale parameter of Weibull probability density function. Operation complexity of EVM is discussed in~\cite{Rudd_EVM}. This classifier is hardware friendly and can run effectively on embedded system platform~\cite{Jaafarzade_END_TO_End}. We used the Python (Pytorch) GPU accelerated EVM library of paper \cite{jafarzadeh2021automatic} available in GitHub repository \cite{EVM_library}.

\section{Spectral Analysis}
\subsection{Spectrum Estimation - Background}
The myoelectric signal processing is applicable for gesture recognition of human hand based on the fact that there is a monotonic relationship between the properties of EMG signal and that of its related finger movement. Various finger gestures have different EMG signal with different properties. To find the pattern recognition in human hand, EMG signal should be scrutinized in real time. However, due to stochastic characteristics of the EMG signal, it is not processed instantaneously. Instead, it is processed in time blocks called sliding windows that can be overlapped or disjoint. Filtering of EMG signal is another inevitable part of signal processing as these signals are contaminated via environmental factors and therefore, the removal of the noise with filtering approaches is required such that valuable information of signal could be preserved. 

Spectral analysis is a signal processing technique to discriminate properties of signals in various applications such as EMG signal and the corresponding finger movements. Based on stochastic nature of EMG signal, it is considered as a random process, and therefore, Fourier series of such condition could not be defined without using autocorrelation model. Autocorrelation model is the way to understand the spectral features of such random processes. This model is described as $R(m) = E[V(n)V(n+m)]$. R correlates the random processes with the $m$ as the lag. Power spectral density is the Fourier transform of R. This transformation involves spectral characteristics of the signal and enables us to extract features in the frequency domain. As R is unknown, the PSD should be gained by estimation. Two different methods have been proposed for the estimation of PSD: non-parametric and parametric.
Nonparametric methods utilize periodogram. The goal of periodogram is to identify innate periodic signals by finding the magnitude of different frequencies. However, the variance of estimation is about the square of its corresponding amount. But in parametric methods, the random process itself is considered as parametric model. The advantage of the latter method is that the amount of parameters for the PSD and models are approximately the same. 

\subsection{ Parametric PSD - Reflection Coefficients}
Let $\epsilon(n)$ denote a white noise with average zero and variance of $\sigma^2$. For the stationary random process $V(n)$, the auto-regressive model is represented as:
\begin{equation}
    \label{equ: AR}
     V(n) = -\sum_{i=1}^{p} a_i V(n-i) + \epsilon(n) ,
\end{equation}
where $a_i$s are the model parameters and $p$ denotes the order of the auto-regressive model.
The model parameters are used to calculate the power spectral density of the stationary random process as:
\begin{equation}
    P(f) = \frac{\sigma^2}{|1+ \sum_{i=1}^{p} a_i \exp(-j2 \pi i f) |^2} ,
\end{equation}
where $i = \sqrt{-1}$.
Different methods can be applied to adjust the model parameters such as Yule-Walker~\cite{stoica1997introduction} method and Burg's method~\cite{agarwal1998automatic}.
In this paper, we will focus on Burg's method to estimate the power spectral density of the input frame.
Burg's method fits an auto-regressive model to the signal by minimizing (least squares) the forward and backward prediction errors. Such minimization occurs with the auto-regressive parameters constrained to satisfy the Levinson-Durbin recursion. Then, without computing the autocorrelation function which is computationally expensive, Burg's method approximates the reflection coefficient directly. In Burg's method, equation (2) is reformulated as a Finite Impulse Response (FIR) filter, i.e.,

\begin{equation}
    \epsilon (n) = \sum_{i=0}^{p} a_i V(n-i),
\end{equation}
where in this configuration $a_0 = 1$.
The $p$-order FIR filter contains $p$ taps and is implemented by cascading $p$ building components.
The lattice structure of a $p$-order digital FIR filter is realized as a $p$-cascaded first-order lattice components. The lattice structure can be represented by scalar reflection coefficients $k_i$ as:
\begin{align}
    \left\{\begin{matrix}
f_i(n) &= f_{i-1} (n) + k_i b_{i-1} (n-1) \\ 
b_i(n) &= b_{i-1} (n) + k_i f_{i-1} (n-1) 
\end{matrix}\right. ,
\end{align}
where $f_i(n)$ and $b_i(n)$ are forward and backward prediction errors at the $i$-th stage, respectively.  
Historically, reflection coefficients are used for evaluating statistical hypothesis corresponding to the auto-regressive model.
Considering $V(n)$ as the input EMG signal, at the first stage we have:
\begin{equation}
    f_0(n) = b_0(n) = V(n). 
\end{equation}
Also, for the the output of the final stage we expect:
\begin{equation}
    f_p(n) = \epsilon(n). 
\end{equation}

For the cases where the auto-regressive model parameters $a_i$s are known, the reflection coefficients $k_i$s are calculated iteratively; Burg's method provides a technique for parameter estimation of the auto-regressive. This technique is based on optimization of forward and backward prediction errors, i.e., $f_i(n)$ and $b_i(n)$ at the $i$-th stage.\\

The model order $p$ is another unknown parameter that must be chosen carefully. Low order models are not able to catch the spetrum peaks and the result will produce a smooth spectra. Also, high order models would be sensitive to the noise and the output will induce false peaks in the spectra. 
Let $K \triangleq [k_1, k_2, \cdots, k_p]$ denote the reflection coefficients of the model by choosing appropriate $p$ empirically. In our work, vector $K$ is considered as the extracted features from input EMG signals to be fed directly into a classifier. 

\begin{table}[t]
  \centering
\caption{PERFORMANCE OF VARIANCE CLASSIFIERS (IN PERCENT) FOR EMG DATASET~\cite{Khushaba_towards}}
\begin{tabular}
{>{\centering\arraybackslash}p{1cm}  >{\centering\arraybackslash}p{1cm} >{\centering\arraybackslash}p{1cm} >{\centering\arraybackslash}p{1cm} >{\centering\arraybackslash}p{1cm} }
  \hline
  \hline
  Classifier & Acc & Prec & Rec &  F1  \\
  \hline
  \hline
 \textbf{EVM} & \textbf{91} & \textbf{91.4} & \textbf{91} & \textbf{90.9}  \\
  \hline
  SVM & 87.5 & 87.9 & 87.5 & 87.6  \\
  \hline
  KNN & 89 & 89.4 & 89 & 88.6  \\
  \hline
  DT & 44 & 48.1 & 44 & 40.6   \\
  \hline
  RF & 62.5 & 65.9 & 62.5 & 62.9  \\
  \hline
  LR & 51 & 52.6 & 51 & 51.1  \\
  \hline
   GNB & 40.5 & 43 & 40.5 & 38.9  \\
  \hline
  XGB & 76 & 77 & 76 & 76.2  \\
  \hline
  \hline
\end{tabular}
\label{test results of classifiers}
\end{table}

\section{Experimental Results and discussion}
\subsection{Set up}
In this paper, the EMG dataset collected at Sydney University~\cite{Khushaba_towards} is used for feature extraction and classification. The reflection coefficients of the autoregressive model is used for feature extraction. Then, we will compare the accuracy results of EVM against seven well known classifications reported in the literature. We have implemented all classifiers (see Table~\ref{test results of classifiers}) in Python and ran them in a windows-based laptop with $3.00$ GHz CPU and $16.0$ GB memory.

\begin{figure*}[t]
    \centering
    \includegraphics[width=0.77\linewidth]{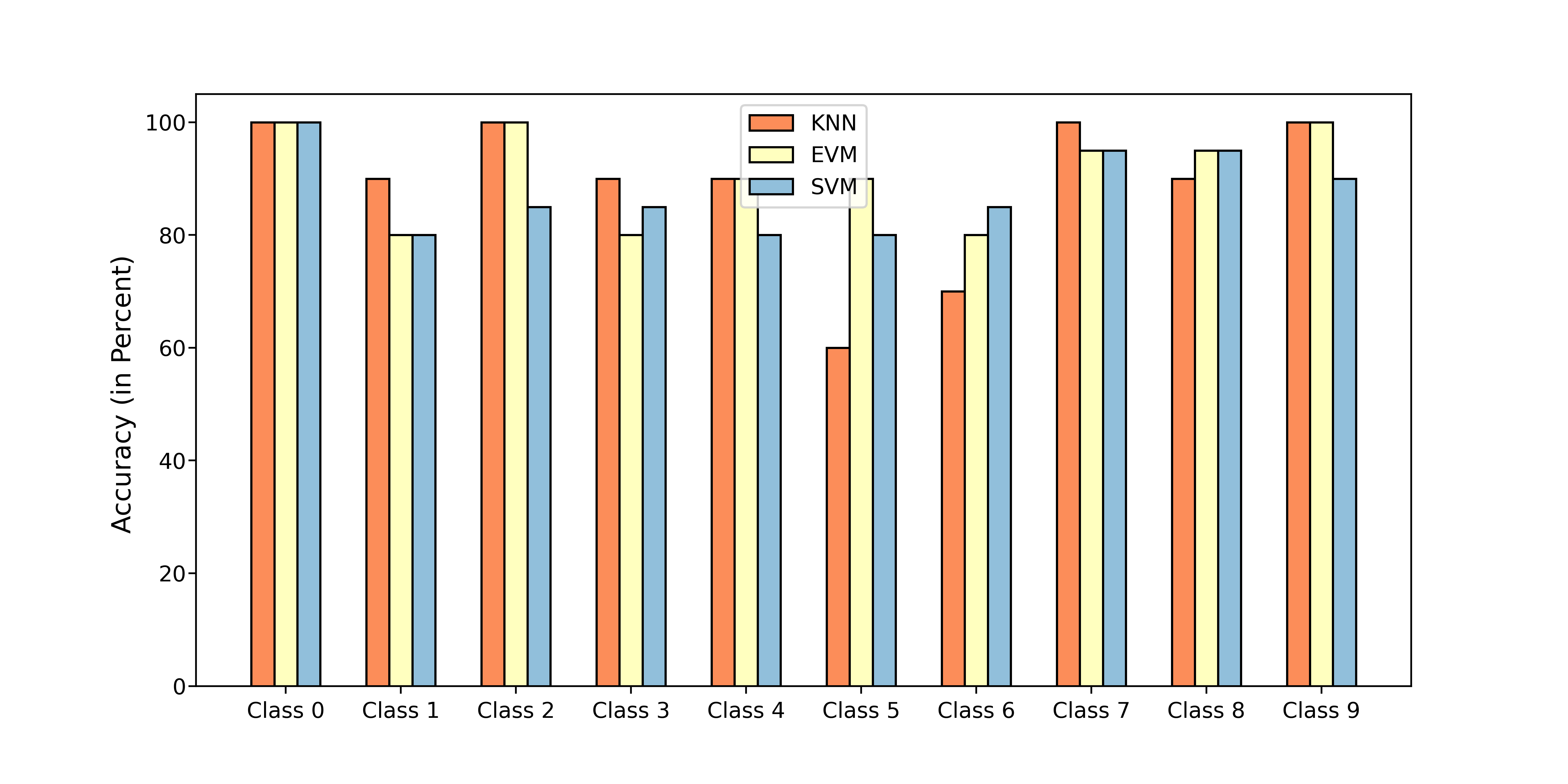}
    \caption{Accuracy of each class for KNN, SVM and EVM}
    \label{class accuracy}
\end{figure*}

\begin{table*}[t]
  \centering
\caption{COMPARING THE RESULTS OF DIFFERENT APPROACHES FOR THE SAME EMG DATASET.}
\begin{tabular}{ >{\centering\arraybackslash}p{1.10cm} >{\centering\arraybackslash}p{3.30cm} >{\centering\arraybackslash}p{2.25cm} >{\centering\arraybackslash}p{1.25cm}>{\centering\arraybackslash}p{1.25cm} }
  \hline
  \hline
Approaches & Features & Classifier &  Clasees & Avg. Acc \\
  \hline
  \hline
 \cite{ariyanto2015finger} & TD+Hjorth+RMS & ANN & 5 & 96.7 \\
  \hline
  \cite{naik2014nonnegative} & AR + RMS & ANN+NMF & 5 & 92 \\
  \hline
  \cite{Khushaba_towards} & TD+AR+Hjorth & KNN+SVM+Fusion & 10 & 90   \\
  \hline
  \cite{Heydarzade} & Reflection Coefficients & SVM & 10 & 89 \\
  \hline
  \textbf{Ours} & \textbf{Reflection Coefficients} & \textbf{EVM} & \textbf{10} & \textbf{91}  \\
  \hline
  \hline
\end{tabular}
\label{comparing the results of accuracy}
\end{table*}

\subsection{Dataset}

The dataset which is used in this paper is provided by Center of Intelligent Mechatronic Systems at the University of Technology at Sydney\cite{Khushaba_towards}. This dataset includes EMG data of eight participants (six men and two women at the range of $25$ to $30$ years old) performing ten different finger movements consisting five individuals as Thumb (T), Index (I), Middle (M), Ring (R), Little (L) and five combined movement as Thumb-Index (T-I), Thumb-Middle (T-M), Thumb-Ring(T-R), Thumb-Little (T-L) and closed hands. In Fig. 2., these fingers movements are illustrated. The duration of each movement from rest pose to contraction pose is five seconds. EMG signals are sensitive to the condition of experiment and effects of neighbor limb. For uniformity, participants sat on the armchair such that their arms were fixed and stable.  The EMG data is recorded with two channels at $4,000$ Hz and then amplified to total gain of $1,000$. Each movement of fingers is repeated six times. We select the first four trials for the training of the classification methods and the last two for the test of the methods in order to check the accuracy of the classifiers. Fig. 3. shows the EMG signal of a combined movement, i.e., Thumb-Index.

\subsection{Results and Discussion}
At the first step, filtering, standard scaling and feature extraction are done. To have a better comparison between classifiers, the hyperparameters of each  classifier are tuned such that achieves the best possible accuracy. 
After determining the hyperparameters for each classifier, for the same dataset the comparison is done. In Table~\ref{test results of classifiers}, four performance metrics (accuracy, precision, recall, and F1 score) are compared. The results indicate that EVM outperforms other conventional classifiers. To the best of our knowledge, our work is the first to utilize EVM for EMG analysis. Extreme value machine is a machine learning technique that models probability of each class of training data by fitting a single or multiple Weibull distribution and is not limited to specific condition~\cite{Rudd_EVM}.  

The EVM has the best performance which makes it a powerful tool for EMG classification. 
The higher accuracy of EVM is obtained from cosine metric which is related to the radially covering of each class with EVM. The tail size  and cover threshold are tuned by $27$  and $0.3$ respectively. Extrem vectors (values) are the EMG features that are located in the tail of distribution of each class. Also, the most of widely used methods do not model the tail of probability distribution of each class. So, we expect that although they have very good performance, but they perform poorly on the tail of distribution. The accuracy for each class is shown in Fig. 4. For KNN, the accuracy for different classes is not consistent. For example, the accuracy of KNN for class five is below $60\%$. SVM has lower fluctuations in the results, but its accuracy is $100\%$ only for the class $0$, while for other classes, its accuracy is low. EVM has robust accuracy for each class and its accuracy is not fluctuating a lot for various classes. 

In Table 2, we compare the accuracy with similar works reported in~\cite{Heydarzade, ariyanto2015finger, naik2014nonnegative, Khushaba_towards}. Authors in~\cite{naik2014nonnegative} categorized ten finger movements into two groups of five movements and extracted AR and Root mean Square (RMS) of amplitude of EMG samples as their features. By running artificial neural network (ANN), they could reach $92\%$ accuracy. EVM is a powerful method constructed based on the distribution of data which is rarely considered on many other well-known classifiers. 

Reference~\cite{ariyanto2015finger} also restrained their research to five finger movements. They used time domain, Hjorth and RMS as their features and their classifier was ANN. Reference~\cite{Khushaba_towards} used a combination of time domain features, Hjorth and AR, and trained either a KNN classifier or a SVM classifier on all ten movements. Also, they implemented the fusion approach to vote among different classification methods to get better results. \cite{Heydarzade} used reflection coefficients of autoregressive model  as the extracted features for classification. Similar to~\cite{Heydarzade} and~\cite{Khushaba_towards}, we concentrate on all ten movements, but our accuracy is higher than the similar works. Additionally, our accuracy is obtained without extra processing mechanism like feature selection, feature reduction, voting or fusion.

\section{Conclusion}
In this paper, we proposed EVM as a promissing classifier for EMG datasets. We analyzed an open source data with eight subjects and ten classes. For feature extraction, we used reflection coefficients. 
These features capture the frequency properties of EMG signals. Then we compared the accuracy of EVM against seven conventional classifiers. The average accuracy of EVM is $91\%$ which is superior or competitive accuracy. 

The main focus of this paper was to utilize extreme value machine and validate its strength on EMG signals like Electroencephalography (EEG), Electrocardiography (ECG), etc. Other aspects of research on extreme value machine such as its application on other biomedical signals, its combination with other neural network architectures such as convolutional neural network, graph convolutional neural network, and recurrent neural network remain open for future studies. 

\bibliographystyle{ieeetr}
\bibliography{Ref.bib}
\end{document}